\documentclass[useAMS,usenatbib]{mn2e}
\usepackage{graphicx}

\def\ms{\mbox{$\rm M_\odot$}}
\def\ds{\mbox{$d_\odot$}}

\def\dgc{\mbox{$R_{GC}$}}

\def\ssfr{\mbox{$\Sigma_{SFR}$}}

\def\ta{\mbox{$t_A$}}

\def\tdis{\mbox{$t_{dis}$}}
\def\kms{\mbox{$\rm km\,s^{-1}$}}
\def\mmy{\mbox{$\rm\ms\,Myr^{-1}$}}
\def\mmk{\mbox{$\rm\ms\,Myr^{-1}\,kpc^{-2}$}}

\title[The SFR in the Solar neighbourhood]{Constraining the star formation rate in
the Solar neighbourhood with star clusters}

\author[C. Bonatto and E. Bica]{C. Bonatto$^1$ and E. Bica$^1$\\
$^1$ Departamento de Astronomia, Universidade Federal do Rio Grande 
do Sul, Av. Bento Gon\c{c}alves 9500\\
Porto Alegre 91501-970, RS, Brazil}

\begin{document}

\pagerange{\pageref{firstpage}--\pageref{lastpage}}

\maketitle

\label{firstpage}

\begin{abstract}
This paper investigates the star formation rate (SFR) in the Solar neighbourhood. First, we 
build the local age distribution function (ADF) with an updated sample of 442 star clusters 
located at less than 1\,kpc from the Sun. Next, we define the SFR, compute the individual 
mass evolution of a population of artificial clusters covering the broad range of parameters 
observed in actual clusters, and assume 100\,\ms\ as the low-mass limit for effective cluster
observation. This leads to a simulated ADF, which is compared to the low-noise Solar 
neighbourhood ADF. The best match corresponds to a non-constant SFR presenting two 
conspicuous excesses for ages $\le9$\,Myr and between $220-600$\,Myr (the local starburst).
The average formation rate is $\overline{SFR}\approx(2500\pm500)\,\mmy$, corresponding to 
the average surface formation rate $\overline{\ssfr}\approx(790\pm160)\,\mmk$. These values
are consistent with the formation rate inferred from embedded clusters (ECs), but much lower 
($\la16\%$) than that implied by field stars. Both the local starburst and the recent star 
formation period require $SFR\sim2\times\overline{SFR}$ to be described. The simulations show 
that $91.2\pm2.7\%$ of the clusters created in the Solar neighbourhood do not survive the first 
10\,Myr, which is consistent with the rate of EC dissolution.
\end{abstract}


\section{Introduction}
\label{Intro}

It is currently accepted that star formation is scale-free and hierarchical, with high-velocity 
turbulent gas forming large-scale structures, while low-velocity compression forms small clumps  
(\citealt{ElmeCon}). A natural consequence of this scenario is that young stellar groupings would 
be hierarchically clustered, with the great star complexes at the largest scales and the OB 
associations and subgroups, clusters and cluster sub-clumps, at the smallest (e.g. \citealt{Efremov95}).  
The bottom line is that star formation appears to occur primarily in clusters. Consequently, star 
clusters appear to be excellent tracers of the star formation rate (SFR) and history of the Galaxy, 
provided that their fundamental parameters - especially the age - are known.

However, star clusters have limited lifetimes. At the earliest phases, most of the embedded star 
clusters (ECs) dissolve into the field on a time-scale of a few $10^7$\,Myr. Dissolution here
occurs mainly because the gravitational potential can be rapidly reduced by the impulsive gas 
removal by supernovae and massive star winds associated with this early period. Thus, a significant
fraction of the stars, especially of low mass, end up moving faster than the scaled-down escape velocity, 
and may escape to the field (e.g. \citealt{GoBa06}). 

Dissolution also affects the open clusters (OCs) that survive the early phase. Even around the 
Solar circle, most OCs dissolve long before reaching an age of $\sim1$\,Gyr (e.g. \citealt{DiskProp}). 
The disruption timescale, in general, scales with mass as $\tdis\sim M^{0.62}$ (e.g. \citealt{LG06}),
which means that Solar neighbourhood clusters with mass in the range $10^2-10^3\,\ms$ dissolve on
a timescale $\rm75\la\tdis(Myr)\la300$. This occurs because OCs continually undergo mass segregation 
and evaporation, tidal interactions with Galactic substructures, shocks with giant molecular clouds 
(GMCs), as well as mass loss due to stellar evolution. By decreasing the total cluster mass - and 
the collective gravitational potential, these processes affect the internal dynamics and accelerate 
the cluster dynamical evolution. Eventually, the majority of the OCs dissolve in the Galactic stellar 
field. 

Thus, the only way for recovering the SFR using the observed star cluster population is by taking
the mass-loss processes affecting clusters of different masses and orbits into account. To 
this effect, semi-analytical descriptions of the main mass-loss processes have become available 
recently (e.g. \citet{Lamers10}; \citealt{KASS07}). In the present paper we employ these process 
to simulate the mass evolution of a population of artificial clusters. Properties of the local 
SFR are then derived by comparing the simulated age distribution function (ADF) with that built 
for clusters in the Solar neighbourhood.

This paper is organised as follows. In Sect.~\ref{OCADF} we build the ADF for the Solar
neighbourhood. In Sect.~\ref{TCADF} we briefly discuss the cluster mass-loss process. In 
Sect.~\ref{sfr} we simulate the observed ADF and use it to constrain the local SFR. Concluding 
remarks are given in Sect.~\ref{Conclu}.

\section{Building the observed cluster ADF}
\label{OCADF}

The number of star clusters with accurate age and distance determinations has been steadily 
increasing over the last years. This is particularly true for the very young clusters, most
of which so embedded in their parent gas and dust cloud that their stellar content is essentially 
inacessible to optical photometry. However, the availability of uniform, wide-field, and rather 
deep near-infrared surveys (e.g. 2MASS\footnote{The Two Micron All Sky Survey, All Sky data 
release (\citealt{2mass06})}, GLIMPSE\footnote{Galactic Legacy Infrared Mid-Plane Survey 
Extraordinaire (\citealt{Benjamin03})}, some covering essentially all the sky (2MASS), has 
led to the discovery - and allowed a robust parameter derivation - of many such embedded clusters 
(ECs - e.g. \citealt{DBSB03}; \citealt{BDSB03}; \citealt{Kumar04}); \citealt{KKC06}); \citealt{N6611};
\citealt{DBSB48}; \citealt{N2244}; \citealt{Pi5}; \citealt{vdB92}), together with some old and/or 
very reddened OCs (e.g. \citealt{Froeb07}; \citealt{OldOC2}; \citealt{OldOC1}; \citealt{LKStuff}; 
\citealt{Froeb10}; \citealt{Teu34}). 

Together with the several hundred objects already indexed in the widely-used star cluster 
databases WEBDA\footnote{\em www.univie.ac.at/webda} and DAML02\footnote{Catalog of Optically 
Visible Open Clusters and Candidates, \citet{DAML02}.}, the recent discoveries (together with 
parameter derivation for poorly-studied and/or unstudied objects) are proving invaluable in 
constructing a more detailed picture of the cluster ADF, especially in the Solar neighbourhood 
and for very young clusters (Sect.~\ref{sfr}). 

We started by searching WEBDA and DAML02 for clusters with available age (\ta) and distance 
from the Sun (\ds). However, given the amount of new data routinely published, it usually takes 
a considerable time for both databases to incorporate the recently discovered clusters, or to
update previous entries with newly derived parameters. Then, we complemented the sample by 
searching the recent literature for clusters that still are not listed in either database. 
Cluster designations and coordinates have been checked among all sources (WEBDA, DAML02, and 
literature) to avoid duplicity. When multiple values of age and/or distance occurred, we adopted 
those based on colour-magnitude diagrams (CMDs), or the more recent. The final sample contains 
1718 clusters (ECs and OCs) with age and distance, of which 442 are closer than 1\,kpc from the 
Sun\footnote{Upon request, the table with the age and distance from the Sun for the 442 nearby
clusters is available from one of us (C.B.).}. By far, most of the parameters have been taken from WEBDA 
and DAML02, which do not provide measurement uncertainties. Thus, based on our experience in working 
with clusters of different ages and distances, we adopted the following uniform error attribution: 
10\% for $\ds<1$\,kpc, 15\% for $\rm1<\ds(kpc)<5$, 20\% for $\rm5<\ds(kpc)<9$, and 25\% for $\ds>9$\,kpc; 
35\% for $\ta<20$\,Myr, 30\% for $\rm20<\ta(Myr)<100$, 20\% for $\rm100<\ta(Myr)<2000$, and 50\% for 
$\ta>2000$\,Myr. 

%
%

Age uncertainties are explicitly incorporated into the ADF, which is defined as the fractional 
number of clusters {\em per} Myr, $ADF\equiv dN/d\ta$. Formally, if measurements of a given
parameter $\chi$ are normally (i.e. Gaussian) distributed around the average $\bar\chi$ with 
a standard deviation $\sigma$, the probability of finding it at a specific value $\chi$ is 
given by $P(\chi)=\frac{1}{\sqrt{2\pi}\sigma}\,e^{{-\frac{1}{2}}\left(\frac{\chi-\bar\chi}
{\sigma}\right)^2}$. To build this ADF we first define a set of bins spanning the whole 
range of ages, and having widths that increase with age (to account for the decreasing 
number of clusters at older ages). Then, for a cluster with age and uncertainty $\ta\pm\sigma$, 
we compute the probability that the age corresponds to a given bin, which is simply 
the difference of the error functions at the bin borders. By doing this for all 
clusters and age bins, we have the number-density of clusters in each age bin. By 
definition, the integral of the ADF over the whole range of ages is the number of 
clusters. Subsequently, bin widths can be adjusted to minimise the errors, so that the 
resulting ADF has statistically meaningful values over all ages. 

\section{Overview of the mass-loss processes}
\label{TCADF}

Star clusters lose mass continually by a combination of processes associated with stellar 
evolution and dynamical interactions (both internal and external to the cluster). Robust 
analytical descriptions of the mass-loss processes - for clusters characterised by a wide
variety of parameters and orbiting in different environments - have become available in 
recent years, with model parameters derived from theoretical grounds (e.g. 
\citealt{Spitzer87}; \citealt{Lamers10}) and N-body simulations (e.g. \citealt{BM03}; 
\citealt{GB08}). Formally, the time-rate of change of mass of a cluster that was formed 
with the mass $M_i = M(0)$ can be expressed as:

\begin{equation}
\label{eq1}
\frac{dM}{dt}=\sum_{p=1}^6\left(\frac{dM}{dt}\right)_{p},
\end{equation}
where the mass-loss process $p$ are: (1) stellar evolution, (2) tidal effects by a steady 
field, (3) shocks with spiral arms, (4) encounters with GMCs, (5) evaporation, and (6) 
ejection. For processes (1) - (4) we adopt the semi-analytical approach of \citet{LG06} and
\citet{Lamers10}. In what follows we assume that masses are always expressed in Solar 
masses (\ms) and time in Myr; also, we write the equations in terms of the remaining-mass 
fraction $\mu\equiv M/M_i$. For clarity, we provide below a brief description of these 
processes.

\subsection{Stellar evolution}
\label{StelEvol}

Since we are dealing essentially with disk clusters, we adopt the Solar-metallicity
approximation given by \citet{LG06}, which is based on single-stellar population GALEV 
models (\citealt{GALEV}) and a Salpeter-like (\citealt{Salp55}) mass function:

$$\left(\frac{d\mu}{dt}\right)_{se} = -\mu\left(\frac{a}{t}\right)10^{[(a-1)\log(qq)+qq^a+b]},$$
with $qq = \log(t)-1.0$, $a=0.255$, and $b=-1.805$. This process is relevant for $t>10$\,Myr.

\subsection{Tidal effects}
\label{TidEff}

We adopt the semi-analytical description of the mass-loss of clusters on a steady-tidal 
field of \citet{Lamers10}, which is expected to apply to a broad range of cluster conditions 
and orbit environments:

$$\left(\frac{d\mu}{dt}\right)_{tidal} = -\frac{\mu^{1-\gamma}}{t_o M_i^\gamma},$$ with 
$t_o = t_R\left(\frac{1-\epsilon}{\bar m^\gamma}\right)\left(\frac{R_G}{8.5kpc}
\right)\left(\frac{v_G}{220\kms}\right)^{-1}$, $t_R=13.3$\,Myr for clusters with an
initial concentration factor of the density King profile $W_0=5$, and $t_R=3.5$\,Myr 
for $W_0=7$; $\gamma=0.65$ for $W_0=5$ and $\gamma=0.8$ for $W_0=7$, thus, both $\gamma$ 
and $t_R$ are not independent parameters, since they can be expressed as a function of 
$W_0$; $\epsilon$ is the cluster orbit eccentricity; the average stellar mass at time $t$ 
is given by the interpolation formula $\bar m(t) = 0.6193-0.0362\tau-0.01481\tau^2+0.0022\tau^3$,
where $\tau=\log(t)$; $R_G$ is the cluster's Galactocentric distance, and $v_G$ is the (assumed constant) 
rotation velocity. In what follows we take $v_G=200\,\kms$ for orbits around the Solar neighbourhood.

\subsection{Shocks with spiral arms}
\label{sp}

The energy gain and the mass loss due to the disruptive effect of spiral-arm passages of 
star clusters with planar and circular orbits around the centres of galaxies has been
thoroughly studied by \citet{GAP07}. In  particular, for the Solar neighbourhood,
\citet{GAP07} and \citet{LG06} find:

$$\left(\frac{d\mu}{dt}\right)_{sp} = -5\times10^{-5}\frac{\mu^{0.3}}{(M_i/10^4\ms)^{0.7}}.$$

\subsection{Encounters with giant molecular clouds}
\label{gmc}

Similarly to the spiral arms, \citet{LG06} estimate the energy gain and mass loss due to 
encounters between clusters and GMCs. Assuming GMC parameters typical of the Solar 
neighbourhood, they find:

$$\left(\frac{d\mu}{dt}\right)_{GMC} = -5\times10^{-4}\frac{\mu^{0.3}}{(M_i/10^4\ms)^{0.7}}.$$
The mass loss due to GMCs is thus 10 times higher than that of the spiral arms.

\subsection{Ejection and evaporation}
\label{evap}

Stars can also escape from a cluster by means of ejection and evaporation. Ejection 
occurs when, after a single, close encounter with another star, a member ends up with 
excess velocity with respect to the escape velocity. Evaporation, on the other hand,
is related to a series of more distant encounters that gradually increase a star's
energy, eventually leading it to escape from the cluster. With the respective time
scales taken from Khalisi, Amaro-Seoane \& Spurzem (2007, and references therein),
the mass loss associated with both processes are, respectively:

$$\left(\frac{d\mu}{dt}\right)_{ej} = -\frac{\bar m(G\mu)^{1/2}}{46(M_i R_{hm}^3)^{1/2}},$$ and  
$$\left(\frac{d\mu}{dt}\right)_{ev} = -\frac{\bar m(G\mu)^{1/2}\ln(\gamma_c N)}{13.8(M_i R_{hm}^3)^{1/2}},$$
where the half-mass radius is given by $R_{hm} = 3.75\left(M/10^4\right)^{0.1}$
(\citealt{Larsen04}), $\gamma_c=0.11$ is the Coulomb factor, $N=M(t)/\bar m(t)$ is the
number of stars at time $t$, and $G$ is the gravitational constant.

\subsection{The adopted procedure}
\label{Proc}

In summary, given a cluster of initial mass $M_i$, and a set of model parameters (\dgc, 
$W_0$, $\gamma$, $\epsilon$, $t_R$), Eq.~\ref{eq1} should be solved to find the mass still 
bound to the cluster at a later time $t$, $M=M(t)$.

\section{Interpreting the local ADF}
\label{sfr}

The Solar neighbourhood ADF has been subject to previous investigation. In particular, \citet{Lamers05} 
and \citet{LG06} employed the cluster dissolution processes 1---4 (Sect.~\ref{TCADF}) to 
study the ADF of 114 OCs closer than $\ds=0.6$\,kpc taken from \citet{Kharchenko05}. Their 
ADF can be reasonably well described by a nearly-constant SFR (in bound clusters of mass
$10^2<M(\ms)<3\times10^4$) of $SFR=400\,\mmy$, corresponding to a surface formation rate
$\ssfr\approx350\,\mmk$, which is considerably lower than the $\ssfr=700-1000\,\mmk$ 
inferred from ECs (\citealt{LL2003}), and the $\ssfr=3000-7000\,\mmk$ from field stars 
(\citealt{MS79}). They also find evidence of a local burst of star formation that took 
place between 250---600\,Myr ago (also present in \citealt{Piskunov06}). 

Besides cluster dissolution, observational completeness is also important for the ADF shape. 
High absorption and crowding, usually associated with fields dominated by disc and bulge stars, 
affect cluster detectability, especially the faint and/or poorly-populated ones. \citet{DiskProp} 
show that most of the intrinsically faint and/or distant clusters must be undetected in the 
field, particularly in bulge/disc directions. They also find that the completeness-safe zone 
might extend up to $\ds\approx1.4$\,kpc inside the Solar circle, but considerably more 
outside.

Thus, to minimise observational completeness effects, we restrict the analysis to the  
region $\ds\le1$\,kpc, which reduces the number of clusters to 442, but still a 
statistically significant sample. Besides, this condition also guarantees that we are 
dealing with clusters subject to somewhat similar physical conditions. By construction 
(Sect.~\ref{OCADF}), the resulting ADF (Fig.~\ref{fig1}) is rather smooth, with small 
errors over all ages, except for $\ta\ga3$\,Gyr, which reflects the scarcity of Gyr-old 
clusters in the region. Also conspicuous are the excesses between $\rm200\la\ta(Myr)\la600$, 
and especially for $\ta\la10$\,Myr (see below). Since we work with a significantly larger 
sample than \citet{Lamers05} and \citet{LG06}, our approach provides a better definition 
of structures in the ADF.

\subsection{ADF simulation}
\label{SimulADF}

Instead of assuming particular values for the several model parameters (\dgc, $W_0$, $\gamma$, 
$t_R$, $\epsilon$, number, age, initial mass, and the orbit of clusters), we adopt a simplifying 
approach. Starting with a pre-defined SFR, we compute the time-evolution of the individual mass 
of star clusters created with parameters covering the observed range in orbits around the Solar 
circle (\citealt{Lamers10}). We assume that the relevant model parameters can take on any value 
within $0.0\le\epsilon\le0.8$ (also allowing for circular orbits), $5\le W_0\le7$, 
$0.65\le\gamma\le0.8$, and $\rm3.5\le t_R(Myr)\le13.3$ (the latter two are expressed as a 
function of $W_0$ as $t_R = 37.8-4.9W_0$ and $\gamma = 0.275+0.075W_0$); also, given the 
restriction $\ds\le1$\,kpc, we have $\rm7.5\le\dgc(kpc)\le9.5$. Each cluster is created (at 
time \ta) with an individual mass in the range $M_{min}<M_i(\ms)<M_{max}$, but following the
number-frequency (or probability) $dN/dM_i\propto M_i^{-2}$ (e.g. \citealt{ElmeCon})\footnote{This 
is done by $M_i=\frac{M_{min}}{1-n(1-M_{min}/M_{max})}$, where $n$ is uniformly distributed 
within [0,1].}. We adopt $M_{min}=10$\,\ms\ and $M_{max}=7.5\times10^4$\,\ms\ as the minimum 
and maximum initial cluster mass, respectively. Then we solve Eq.~\ref{eq1} to find the mass 
remaining in each cluster after evolving for a time from $t=0$ to $t=\ta$, and build the 
simulated ADF. Similarly to \citet{LG06}, we assume 100\,\ms\ as the minimum cluster mass 
for completeness not being an issue to detectability, i.e, only clusters more massive than 
100\,\ms\ are accounted for in the ADF. According to this approach, the number of free 
parameters reduces essentially to the SFR shape.

The following steps are taken: {\em (i)} We assume that all star formation occurs in clusters, 
define the SFR and compute the total mass $\left(M_{tot}=\int{SFR(t)}\,dt\right)$ converted into 
stars over the Galaxy age; in the case of a non-constant SFR, this should be done separately for 
each SFR segment (or period of time). {\em (ii)} We assign the initial mass $M_i$ (see above)
for one cluster. {\em (iii)} Next, we randomly assign the cluster age \ta\ (it should be 
within the corresponding SFR time segment), and the model parameters. {\em (iv)} Steps {\em (ii)} 
and {\em (iii)} are repeated for more clusters, until the sum of initial masses equals $M_{tot}$. 
{\em (v)} Then we solve Eq.~\ref{eq1} for all clusters from $t=0$ to \ta, and keep only those with 
$M(\ta)>100$\,\ms. {\em (vi)} Finally, stochastic effects are minimised by repeating steps {\em (ii)} 
to {\em (v)} several times ($N_{sim}$). In the present case we use $N_{sim}=100$, and solve 
Eq.~\ref{eq1} for $M=M(t)$ by means of a $4^{th}$ order Runge-Kutta method, with a fixed time step 
of 1\,Myr. 

\subsection{Constraining the local SFR}
\label{SNADF}

\begin{figure*}
\resizebox{\hsize}{!}{\includegraphics{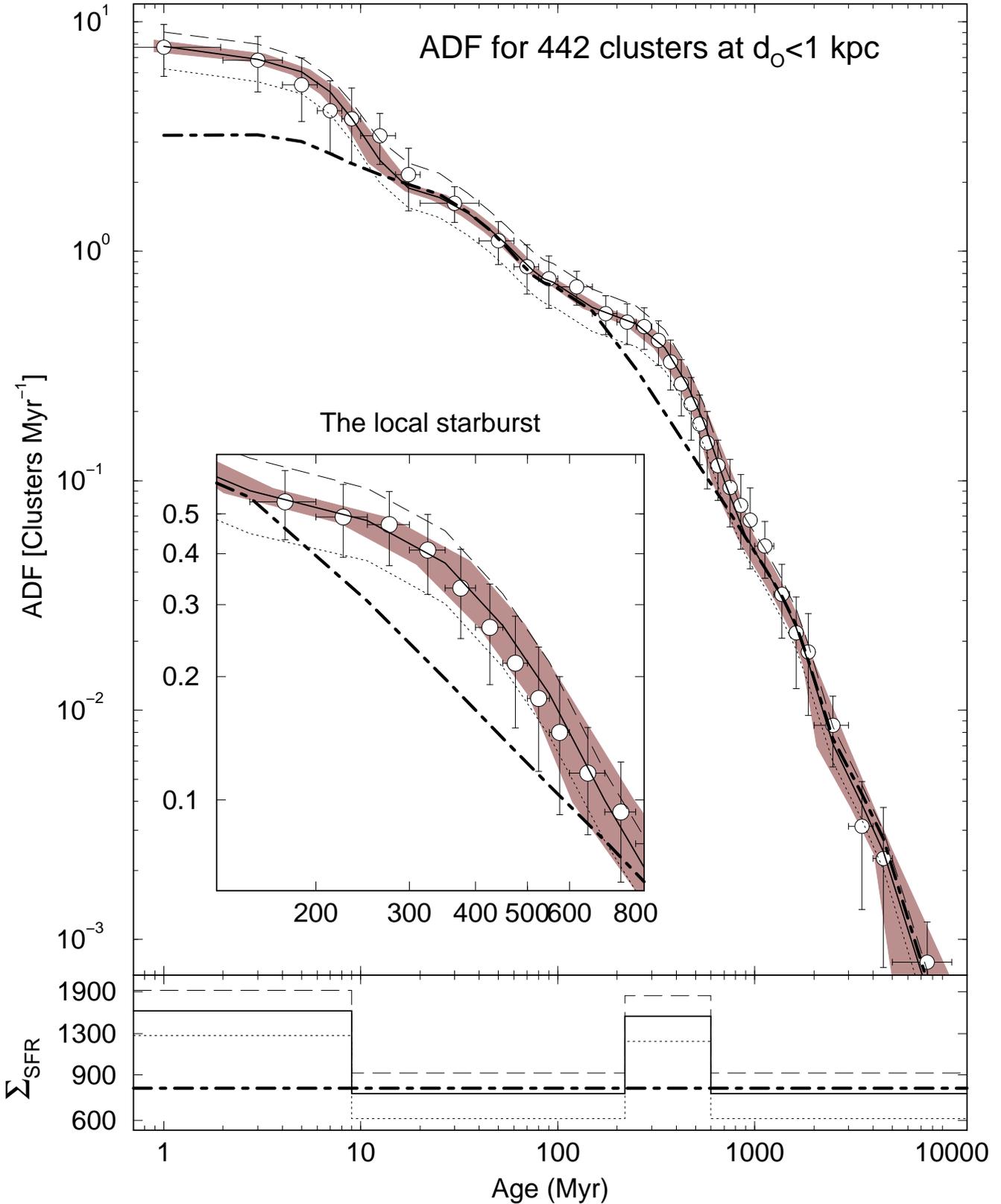}}
\caption[]{Top: The Solar neighbourhood ADF (circles) is well reproduced by a segmented SFR 
(heavy-solid line) with enhanced rates at $\le9$\,Myr and $220-600$\,Myr and small Stochastic 
fluctuations over 100 simulations (shaded region). Also shown are the ADFs produced by \ssfr's 
20\% higher (dashed line) and lower (dotted) than above, together with a constant \ssfr\ (heavy 
dot-dashed line). Inset: Blow-up of the local starburst period. Bottom: The \ssfr's (in
$\rm\ms\,Myr^{-1}\,kpc^{-2}$) used in the top panel.}
\label{fig1}
\end{figure*}

Given the conspicuous excesses (Fig.~\ref{fig1}), it is clear that the Solar neighbourhood 
cluster ADF cannot result from a constant SFR. Instead, the best match has been obtained with 
the segmented SFR: $SFR=5040\,\mmy$ ($\ssfr\approx1600\,\mmk$) for $\ta\le9$\,Myr, $SFR=4800\,\mmy$ 
($\ssfr\approx1500\,\mmk$) for $\rm220\le\ta(Myr)\le600$, and $SFR=2400\,\mmy$ ($\ssfr\approx760\,\mmk$) 
elsewhere (bottom panel of Fig.~\ref{fig1}). Integrating the segmented-SFR ADF
over time yields the same number of clusters as that of the observed ADF. Also, the $1\sigma$ 
boundaries of the adopted solution (shaded region in Fig.~\ref{fig1}), obtained after 100 
different simulations, are very narrow over most of the cluster ages. To check how constrained 
is this solution, we apply the same procedure to SFRs that are 20\% higher and lower than the 
adopted SFR over all time periods. Overall, both SFRs provide a much poorer description of the 
observed ADF than the segmented SFR, especially for clusters younger than $\sim300$\,Myr. Thus, we 
adopt 20\% as the SFR uncertainty. As expected, the ADF produced by a constant SFR ($2500\,\mmy$) 
fails to describe the observed excesses. 

Considering the above arguments, the average values (over the full age range) of the SFR and 
surface SFR are $\overline{SFR}\approx(2500\pm500)\,\mmy$ and $\overline{\ssfr}\approx(790\pm160)\,\mmk$. 
Given the higher number of clusters in our sample, we find a \ssfr\ more than twice that of 
\citet{Lamers05} and \citet{LG06}, considerably more consistent with that inferred from ECs by
\citet{LL2003}, but still significantly lower ($\sim16\%$) than the rate implied by field stars 
(\citealt{MS79}). We also compute the fraction of clusters that do not survive the first 10\,Myr,
$91.2\pm2.7\%$, which agrees with the rate of cluster dissolution after the embedded phase (e.g.
\citealt{LL2003}).

On average, $\sim3.3\times10^5$ clusters are created (at $\ds<1$\,kpc) in each simulation 
with the segmented SFR. This implies that less than 0.15\% of the clusters ever formed in 
the Solar neighbourhood can be observed (i.e, having mass $\ga100\,\ms$). And, if the same 
SFR applies to $\dgc=7.5-9.5$\,kpc, $\sim10^7$ clusters have been created in this region 
since Galaxy formation. 

\section{Summary and conclusions}
\label{Conclu}

In this paper we analyse properties of the Solar neighbourhood SFR, by means of 
a statistically significant ADF built with 442 star (embedded and open) clusters 
closer than 1\,kpc from the Sun. By adopting a simplifying approach, in which the 
mass evolution of artificial clusters is followed over time, we reduce the problem 
to essentially finding the SFR. 

The artificial clusters embody parameters and conditions expected to apply to most 
actual star clusters orbiting not far from the Solar circle. To simulate the observed 
ADF, we employ semi-analytical descriptions of the mass-loss process responsible for 
cluster dissolution, and assume that only clusters with a present-day mass above 
100\,\ms\ can be effectively observed (i.e., take part in the ADF).

The best match between observed and simulated ADFs corresponds to a non-constant SFR,
with enhanced rates for ages the $\le9$\,Myr and $220-600$\,Myr (the so-called local
starburst). The average rate is $\overline{SFR}\approx(2500\pm500)\,\mmy$, corresponding 
to the average density $\overline{\ssfr}\approx(790\pm160)\,\mmk$. These values agree with 
the formation rate inferred from ECs, but represent only $\sim16\%$ of the rate implied
by field stars. Both the local starburst and the recent formation ($\le9$\,Myr) have ADF 
amplitudes suggesting periods with a SFR about twice the average value. We also find that
$91.2\pm2.7\%$ of the clusters created in the Solar neighbourhood dissolve before 10\,Myr, 
which is consistent with the rate of EC dissolution.

\section*{Acknowledgements}
We thank the comments of the referee, Simon Portegies Zwart.
We acknowledge support from the Brazilian Institution CNPq. This publication makes 
use of the WEBDA database, operated at the Institute for Astronomy of the University 
of Vienna.

\label{lastpage}
\end{document}